\documentclass[12pt,noshowpacs,nofootinbib,notitlepage,amsmath,amsfonts]{revtex4-2}
\usepackage{setspace}
\usepackage{graphicx,color}
\usepackage[colorlinks=true,citecolor=blue,linkcolor=blue,urlcolor=blue]{hyperref}
\usepackage{multirow}
\usepackage{enumerate}
\usepackage{array}
\usepackage{booktabs}

\begin{document}  
\title{\color{blue}\Large Negative mass singularities mimicking dark energy}

\author{Bob Holdom}
\email{bob.holdom@utoronto.ca}
\affiliation{Department of Physics, University of Toronto, Toronto, Ontario, Canada  M5S 1A7}

\begin{abstract}
One or two negative mass singularities are found to occur in static inhomogeneous spatially closed solutions to the Einstein equations. The singularities produce a positive Komar mass, and this decreases the size of the cosmological constant relative to normal matter. The energy density of a perfect fluid vanishes at the singularities and is finite elsewhere. Numerical examples of these static solutions are provided, and their stability properties are found to be similar to the Einstein static universe. In an expanding universe, the effect of the singularities is to push the acceleration towards more positive values. Given the sentiment that naked singularities are to be avoided, we review just how benign the negative mass singularity is.
\end{abstract}

\maketitle

The well known example of a static spatially closed universe is the Einstein static universe (ESU) \cite{Ein,ORaifeartaigh:2017uct}.\footnote{Other examples include the flat closed universe with the spatial topology of a torus $T^3$ and a much less trivial construction based on a Sasakian space with two counter-streams of dust \cite{Ishihara:2021gty}.} The ESU has a constant density perfect fluid with $p=w\rho$ and a cosmological constant $\Lambda=4\pi(\rho+3p)$. We shall consider an inhomogeneous generalisation of this spacetime where $\rho$ varies while maintaining isotropy around a special point and its antipodal point. Our results indicate that such a solution to the Einstein equations necessarily has a negative mass singularity at one or both of these points.

The metric that proves to be convenient for describing an inhomogeneous generalisation of the ESU is
\begin{align}
    ds^2=B(x)^2(-dt^2+dx^2)+R(x)^2 d\Omega^2
.\label{e0}\end{align}
$x$ has a finite range $0\leq x \leq L$, and $R(0)=R(L)=0$, so that $x=0$ and $x=L$ are the two poles of the closed space. The particular choice $B(x)=1$ and $R(x)=R\sin(x/R)$ with $L/R=\pi$ describes the isotropic and homogeneous ESU. For our inhomogeneous generalisation, $B(x)$ varies and $R(x)$ still has a maximum somewhere between its vanishing at the two poles. The more familiar form of a general spherically symmetric metric cannot conveniently describe these closed spacetimes.

In Section \ref{s1}, we shall discuss the singularity and its properties, leaving the discussion of complete spacetime solutions to later sections. We focus on static inhomogeneous closed solutions, first in Section \ref{s2} where we show how singularities play a role similar to dark energy, and in Sections \ref{s3} and \ref{s4} where we give numerical solutions and a stability analysis, respectively. We consider an expanding inhomogeneous closed spacetime in Section \ref{s5}, again showing how singularities produce accelerated expansion, and suggesting that the singularities are primordial in origin. Developing further contact with observational cosmology is beyond the scope of this work.

We note that a spacetime with a negative mass singularity does not, by definition, enjoy global hyperbolicity. Global hyperbolicity is a sufficient condition for a well posed PDE theory and a clean definition of QFT, but it is not necessary. The negative mass singularity is known to show that it is not necessary, as we shall review in Section \ref{s1} for the scalar field equation. We shall find a further example of well behaved wave dynamics when we study gravitational perturbations in Section \ref{s4}. Although we find a general lack of problems caused by naked singularities in our analysis, we acknowledge that the cosmic censorship conjecture is a widely held view in general relativity. In particular, this has led to studies of non-singular negative mass configurations \cite{Belletete:2013nqa,Mbarek:2014ppa,Johnson:2019tgc}.

\section{The singularity}\label{s1}
For now, we consider a singularity located at $x=0$. The numerical analysis in Section \ref{s3} shows that close to the singularity we have
\begin{align}
    R(x)\to\alpha x^\frac{1}{2},\quad B(x)\to\beta  x^{-\frac{1}{4}},\quad x\to0
.\label{e16}\end{align}
Energy-momentum conservation $\partial_\mu T^{\mu\nu}=0$ for a perfect fluid with pressure related to energy density as $p(x)=w\rho(x)$ gives
\begin{align}
    \rho(x)=\frac{\rho_1}{B(x)^\frac{1+w}{w}}
.\label{e18}\end{align}
Henceforth, we shall assume that $w>0$, in which case we see that the energy density vanishes at the singularity. The energy density is positive and finite throughout spacetime, as are the curvature invariants $R$ and $R_{\mu\nu}R^{\mu\nu}$ determined by the Einstein equations in terms of $\rho(x)$ and positive $\Lambda$. 

The metric close to the singularity can be transformed to
\begin{align}
    &ds^2 = -\frac{2m}{r} d\tilde{t}^2 + \frac{r}{2m} dr^2 + r^2 d\Omega^2,\label{e19}\\
    &\mbox{with}\quad r = \alpha x^\frac{1}{2},\quad m = \frac{\alpha^{3}}{8\beta^2},\quad \tilde{t} = \frac{2\beta^2}{\alpha} \, t\label{e23}
.\end{align}
This is the near singularity limit of the negative mass Schwarzschild solution. We let $m>0$ be the negative of the negative mass. From this we know the singularity in the Weyl curvature scalar,
\begin{align}
    C_{\alpha\beta\mu\nu} C^{\alpha\beta\mu\nu}\to\frac{48m^2}{r^6}=\frac{3}{4\beta^4}\frac{1}{x^3},\quad x,r\to0
.\label{e17}\end{align}

We may expect that there are power corrections to (\ref{e16}) that are dictated by the negative mass Schwarzschild metric. This metric is not relevant at arbitrarily large $x$ because our solutions are neither vacuum nor asymptotically flat, but for some intermediate $x$ these higher power corrections from the vacuum solution may help provide a more accurate description. We can find these corrections by finding the transformation from the negative mass Schwarzschild solution that continues to produce a metric of the form (\ref{e0}), order by order. The result at a finite order brings us closer to the negative mass Schwarzschild solution,
\begin{align}
    R(x)&=\alpha x^\frac{1}{2} (1+\frac{4\beta^2}{3\alpha^2}x^\frac{1}{2}+\frac{4\beta^4}{9\alpha^4}x+\dots),\nonumber\\
    B(x)&=\beta x^{-\frac{1}{4}} (1+\frac{4\beta^2}{3\alpha^2}x^\frac{1}{2}-\frac{2\beta^4}{9\alpha^4}x+\dots).
\label{e21}\end{align}
The required transformation is as in (\ref{e23}) with $r=R(x)$ replacing $r = \alpha x^\frac{1}{2}$. We shall find good evidence for these corrections from the numerical solutions in Section \ref{s3}. Only higher order terms not shown in (\ref{e21}) would be affected if we had instead started with the negative mass Schwarzschild-de Sitter metric. The effect of normal matter is also negligible at sufficiently small $x$ since $\rho(0)=0$.

A naked timelike singularity can raise issues for the description of normal matter, both for particle-like and wave-like behaviours, but these issues depend on the singularity being considered. It is instructive to review some standard analyses to see how benign the negative-mass singularity actually is.

\subsection{Particle motion} Geodesic behaviour is central to the very definition of a singularity \cite{HawkingEllis,Wald:1984rg}. In the $\theta=\pi/2$ plane close to the singularity where (\ref{e16}) can be used, the geodesic equations reduce to
\begin{align}
    t'(\tau)&=\frac{E}{\beta^2}x(\tau)^\frac{1}{2},\quad \phi'(\tau)=\frac{L}{\alpha^2}\frac{1}{x(\tau)},\\
    x''(\tau)&=\frac{E^2}{4\beta^4}+\frac{L^2}{2\alpha^2\beta^2}x(\tau)^{-\frac{3}{2}}+\frac{x'(\tau)^2}{4x(\tau)},\label{e15}\\
    x'(\tau)^2&=\frac{E^2}{\beta^4}x(\tau)-\frac{L^2}{\alpha^2\beta^2}x(\tau)^{-\frac{1}{2}}-\frac{\kappa}{\beta^2} x(\tau)^\frac{1}{2}.
\label{e2}\end{align}
The last equation can be obtained from $u_\mu u^\mu=-\kappa$ where $u^\mu=(t'(\tau),x'(\tau),0,\phi'(\tau))$ and $\kappa=(1,0,-1)$ define timelike, null, and spacelike geodesics, respectively.

At the point on the trajectory where $x'(\tau)=0$, labelled by $x_{\rm min}$, (\ref{e2}) and (\ref{e15}) become
\begin{align}
    E^2&=\beta^2\frac{L^2/\alpha^2+\kappa x_{\rm min}}{x_{\rm min}^\frac{3}{2}},\label{e1}\\
    x''(\tau)&=\frac{3L^2/\alpha^2+\kappa x_{\rm min}}{4\beta^2x_{\rm min}^\frac{3}{2}}
.\label{e3}\end{align}
If $L>0$ then (\ref{e3}) shows that there is always an outward acceleration at such a point on timelike or null geodesics. First, this shows that there are no circular orbits. Second, it shows that $x_{\rm min}$ as determined by (\ref{e1}) is the distance of closest approach to the singularity on a geodesic with given $E$, $L$, and $\kappa$. $L>0$ geodesics do not reach the singularity, and timelike and $L=0$ (meaning purely radial) geodesics also do not reach the singularity. In the latter case $x_{\rm min}=\beta^4/ E^4$. Using $r = \alpha \sqrt{x}$, this is equivalent to $r_\textrm{min}=2m/\tilde{E}^2$ where $\tilde{E}=\frac{\alpha}{2\beta^2}E$ satisfies $Et=\tilde{E} \tilde{t}$. The singularity acts as a repulsive potential barrier such that no massive particle, no part of any extended object, and no non-radial light reaches the singularity. 

Purely radial light can reach the singularity, albeit with infinite redshift. (\ref{e2}) gives $x'(\tau)^2=(E/\beta^2)^2x(\tau)$, an equation without pathological behaviour at the singularity $x(\tau)=0$. This equation also ensures that $x(\tau)$ remains nonnegative. The solution for a null and inwardly directed radial geodesic is
\begin{align}
    x(\tau)=(\frac{E}{2\beta^2})^2(\tau-\tau_0)^2
.\end{align}
When starting from $\tau<\tau_0$ the radial velocity $x'(\tau)=2(\frac{E}{2\beta^2})^2(\tau-\tau_0)$ changes sign from inward (negative) to outward. The corresponding geodesic equation for the metric (\ref{e19}) is  $r'(\tau)^2=\tilde{E}^2$ and the corresponding solution is $r(\tau)=\tilde{E}|\tau-\tau_0|$. In this case the radial velocity changes sign without vanishing. Both results are consistent with the geodesic being smoothly extended through the singularity while keeping the same physical direction. Although the geodesic equations and their solutions are well behaved, the geodesic reaches a curvature singularity at a finite $\tau$, so it is labelled incomplete, e.g.~\cite{Vargas-Serdio:2020qpu}.

Turning now to the tidal acceleration, this is clearly finite for all geodesics that do not reach the singularity. For the null radial geodesic, it is given by $D^2 \xi^\mu / d\tau^2 = - R^\mu_{\nu\rho\sigma} u^\nu \xi^\rho u^\sigma$ with $\tau$ an affine parameter and with $u^\mu=(E/\beta^2)\sqrt{x}(1,\pm 1,0,0)$ from above. The separation vector $\xi^\mu$ lies in a 2-dimensional transverse space \cite{Wald:1984rg}, the $\theta$ and $\phi$ directions. The null radial vectors $u^\mu$ are the principal null vectors for the Weyl tensor and this means that the contribution from the Weyl tensor vanishes \cite{Wald:1984rg}. The remaining contribution to $R^\theta_{\nu\theta\sigma} u^\nu u^\sigma$ or $R^\phi_{\nu\phi\sigma} u^\nu u^\sigma$ from the Ricci tensor component is finite, and the norm $X(\tau)$ of $\xi^\mu$ satisfies
\begin{align}
    \frac{1}{X}\frac{d^2 X}{d\tau^2}=-4\pi(1+w)\frac{E^2}{\beta^4}\frac{\rho(x)}{B(x)^2}
.\end{align}
Both $\rho(x)$ and $1/B(x)^2$ vanish as $x\to0$, giving $d^2 X/d\tau^2=0$ in this limit. The relevant solution for the separation between different null radial geodesics is $X\sim|\tau-\tau_0|\sim\sqrt{x}\sim r$. Thus, radial light acts as though the singularity is a purely geometrical shear-free and transparent focal point. 

\subsection{Wave motion} More relevant to particles in the quantum world is the behaviour of waves. The massless scalar field equation $\Box \phi=0$ using the metric (\ref{e0}) is
\begin{align}
\partial_t^2\psi_l=\partial_x^2\psi_l+\frac{2R'(x)}{R(x)}\partial_x \psi_l-\frac{B(x)^2}{R(x)^2}l(l+1)\psi_l\equiv A_l\psi_l
,\label{e22}\end{align}
with angular variables separated as $\phi=\sum_{lm}\psi_l(t,x)Y_{lm}(\theta,\phi)$. The question of whether there is unique unitary evolution in the presence of the singularity is usually expressed as to whether $A_l$ is an essentially self-adjoint (ESA) operator. $A_l$ is ESA if only one of the possible behaviours near the singularity, as allowed by $A_l\psi_l=0$, has a finite norm. This in turn requires that the choice of inner product be obvious or unique. This is not quite the case, since different inner products have been proposed depending on the context. One context for the scalar field equation is single-particle relativistic quantum mechanics, where the standard Dirac inner product is used \cite{Horowitz:1995gi,Helliwell:2013qxh}. The other context is quantum field theory, where a conserved inner product related to energy, the Sobelev inner product, is used such that finite norm ensures finite energy \cite{Ishibashi:1999vw}. In QFT the normal modes of the field equation are used in the expansion of a quantum field, and these normal modes must have finite energy. An independent early proposal \cite{Wald:1980jn} for a self-adjoint extension to define time evolution was the Friedrichs extension \cite{ReedSimonI}, and later the energy inner product was used in the demonstration of uniqueness \cite{Ishibashi:2003jd}.

For the case of the negative mass singularity, the question of whether the corresponding $A_l$ is ESA has been considered in both contexts. The conclusions differ; $A_l$ is not ESA in relativistic quantum mechanics \cite{Horowitz:1995gi} and is ESA in quantum field theory \cite{Ishibashi:1999vw}. Here we note that relativistic single-particle QM has well known limitations, limitations that are resolved by QFT. For this reason, it is not difficult to argue that $A_l$ for the negative mass spacetime \textit{is} ESA in the sense that matters most. A somewhat different argument in \cite{Giveon:2004yg} leads to the same conclusion.

Returning to (\ref{e22}) and using (\ref{e16}), this equation gives two possible $x$ behaviours for any $l$,
\begin{align}
    \psi_l\sim c_1+c_2\log(x)\quad\textrm{as}\;x\to0
.\end{align}
The contribution to the energy that is most sensitive to the singularity comes from the spatial part of the kinetic term in the action. The contribution from $c_2$ is proportional to
\begin{align}
    \int_0 dx \sqrt{-g} g^{xx}(\partial_x \phi)^2\sim\int_0 dx \sqrt{x}\cdot\sqrt{x}\cdot (1/x)^2 
\end{align}
and is divergent, while the contribution from $c_1$ is finite. In summary, finite energy waves enjoy unique time evolution in the presence of a negative mass singularity.

\section{Mimicking dark energy (static spacetime)}\label{s2}
The field equations may be written in terms of the spatial curvature ${}^{(3)}\!R$ and the Laplacian $\Delta$ of the spatial part of the metric $\tilde{g}_{ij}$,
\begin{align}
{}^{(3)}\!R(x)&=16\pi\rho(x)+2\Lambda,\label{e10}\\
\Delta B(x)&=\left[4\pi(\rho(x)+3p(x))-\Lambda\right]B(x).\label{e11}
\end{align}
Taking the spatial volume integral of the LHS of (\ref{e11}) and using the divergence theorem, gives $4\pi M_K$, where the Komar mass for a static spacetime is
\begin{align}
    M_K=\frac{1}{4\pi}\int_{\partial V} n^i \partial_i B dA
.\end{align}
$n^i$ with $n^i n_i=1$ is the outward-pointing normal to the boundary $\partial V$ of the volume. For a closed regular spacetime there is no boundary, and the Komar mass vanishes. But our closed spacetime has at least one singularity, so we should excise a small ball around a singularity of radius $x$ and check the contribution of this small ball to the Komar mass. Here, $n^i$ points to the centre of the ball. We find $M_K=\lim_{x\to0}[-R(x)^2B'(x)/B(x)]=\frac{1}{4}\alpha^2$. The corresponding result for metric (\ref{e19}) is $\tilde{M}_{K}=\frac{\alpha}{2\beta^2}M_K=m$.

For a number of such singularities, one or two in our case, the total Komar mass is $M_K^\textrm{total}=\frac{1}{4}\sum_i\alpha_i^2$. $M_K^\textrm{total}$ is $1/4\pi$ times the spatial volume integral of the RHS of (\ref{e11}),
\begin{align}
    M_K^\textrm{total}=\int_0^L dx B^2R^2\left[4\pi(\rho+3p)-\Lambda\right].
\label{e12}\end{align}
We see that it is the sum of $\alpha_i^2$ that has meaning here, and not the sum of $m_i$. We also see that $4\pi(\rho+3p)$ averaged in this way becomes larger than $\Lambda$.

From (\ref{e11}) we may also consider the volume integral of $\Delta B(x)/B(x)=4\pi(\rho(x)+3p(x))-\Lambda$. Integration by parts of the LHS also produces a surface term, but with the extra factor of $1/B(x)$, this surface term vanishes as $x\to0$. The remaining term is the volume integral of $(\partial^i B \partial_i B)/B^2$, where the latter could be written as $a_i a^i$ where $a_i$ is the proper 4-acceleration experienced by static observers. More explicitly, it is $B'(x)^2/B(x)^4$ and so we have
\begin{align}
    \int_0^L dx \frac{R^2}{B^3}(B')^2=\int_0^L dx BR^2\left[4\pi(\rho+3p)-\Lambda\right].
\label{e13}\end{align}
The LHS integral is finite and positive due to inhomogeneity, a nonvanishing $B'(x)$. However, note that any inhomogeneity in these solutions is tied to the presence of singularities. Thus, the effect of singularities as implied by (\ref{e13}) is similar to that of (\ref{e12}). The difference is that (\ref{e13}) provides a spatial volume average of $\rho+3p$ rather than of $(\rho+3p)B$.

We can focus on $4\pi\langle \rho+3p\rangle/\Lambda$ increasing from unity, where the angle brackets denote the volume average, to measure the departure from the ESU. This increase corresponds to a reduction of the cosmological constant relative to normal matter. For our spacetimes with one or two singularities, we shall see how large this ratio can get in the next section.

\section{Numerical solutions}\label{s3}
We may now finally show how the negative mass singularity emerges when the Einstein static universe is altered. We simply choose the energy density $\rho_0\equiv\rho(0)$ such that $4\pi(\rho_0+3p_0)\neq\Lambda$, while maintaining a regular solution at this point, $\lim_{x\to0}[B(x),R(x)]=[1,x]$ for the metric in (\ref{e0}). Then the Einstein equations numerically determine the functions $B(x)$ and $R(x)$. With $B(x)=e^{\nu(x)}$ and $R(x)=e^{\lambda(x)}$, the $tt$ and $xx$ Einstein equations are
\begin{align}
    -2 \lambda ''(x)+2 \lambda '(x) \nu '(x)-3 \lambda '(x)^2+e^{2 \nu (x)-2 \lambda (x)}-\Lambda e^{2 \nu (x)}-8 \pi  \rho  e^{2 \nu (x)}&=0,\\
    2 \lambda '(x) \nu '(x)+\lambda '(x)^2-e^{2 \nu (x)-2 \lambda (x)}+\Lambda e^{2 \nu (x)}-8 \pi  w\rho e^{2 \nu (x)}&=0
.\end{align}
We start slightly away from the origin at $x_0\ll1$ with initial conditions $\nu(x_0)=0$, $\lambda(x_0)=\log(x_0)$, $\lambda'(x_0)=1/x_0$. For a wide range of $\rho_0$ for a given $\Lambda$, we find that $R(x)$, after passing a maximum, will reach a point $x_{\rm max}$, the antipode, where $R(x_{\rm max})=0$, as it does for the ESU. But unlike the ESU, the spacetime is inhomogeneous, and the antipode will have the singularity. The energy density $\rho(x)$ remains positive and finite, except for vanishing at the singularity, and the perfect fluid satisfies all standard energy conditions.

We set $\Lambda=10^{-9}$ in Planck units and start with $w=1/3$. For Fig.~1a we take the initial energy density to be half the ESU value, $4\pi(\rho_0+3p_0)=\frac{1}{2}\Lambda$. Close to the singularity, we find that $R(x)\to\alpha(x_{\rm max}-x)^{\frac{1}{2}}$ and $B(x)\to\beta(x_{\rm max}-x)^{-\frac{1}{4}}$, allowing us to find the values of $\alpha$ and $\beta$. These are given in the caption, along with the ratio $4\pi\langle \rho+3p\rangle/\Lambda$. Fig.~1b has the initial energy density twice the ESU value. Here, the energy-density profile is dramatically different, with the matter concentrated near the singularity. In Fig.~1c and 1d the only change is $w=1/10$. Now it is Fig.~1c that has matter concentrated near the singularity.
\begin{figure}[ht]
\centering
\includegraphics[width=\linewidth]{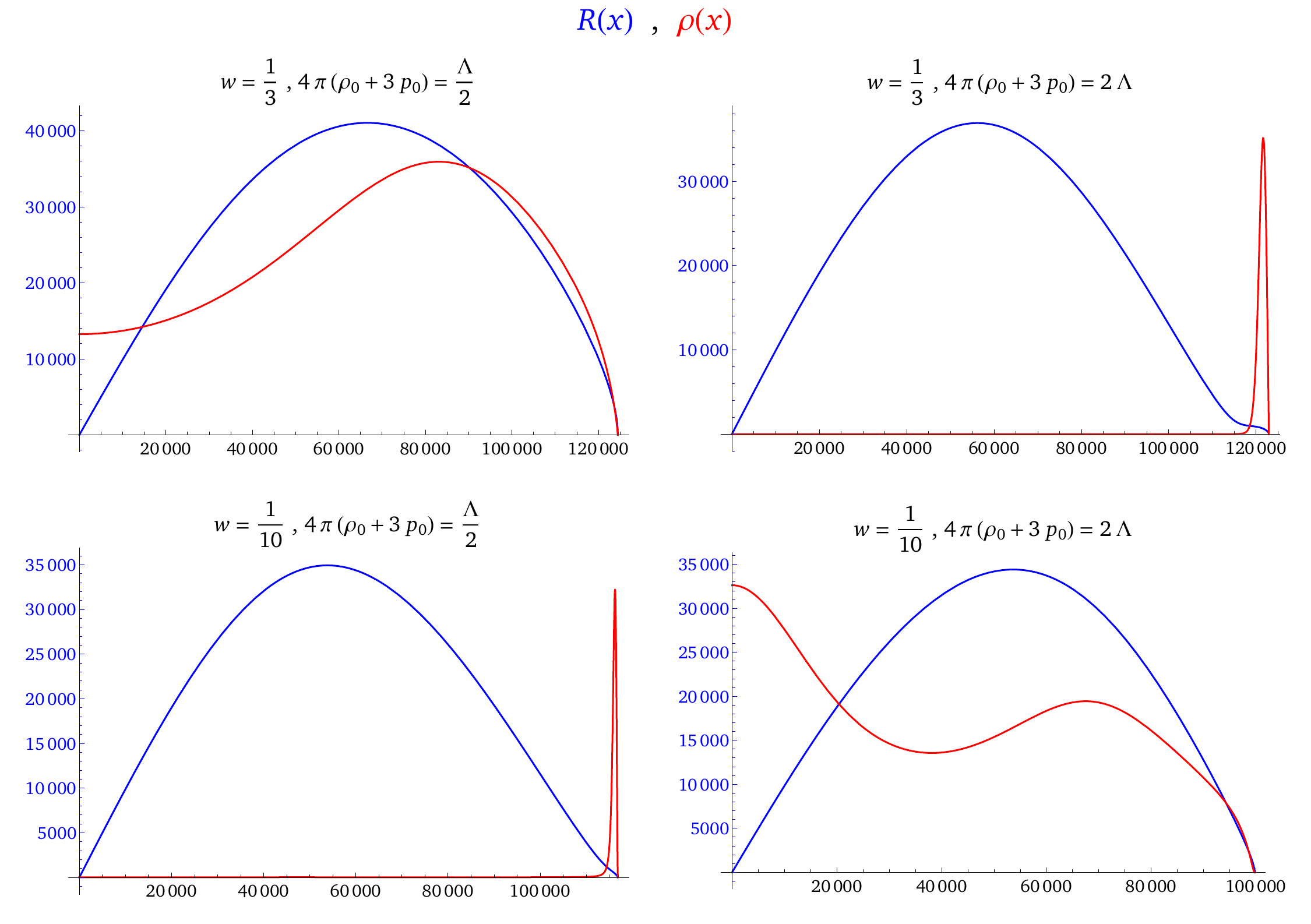}
\caption{The one-singularity solutions with various input values of $w$ and $\rho_0+3p_0$ shown on each plot, with plots a) and b) on first row and c) and d) on second row. The resulting values of ($\alpha$, $\beta$) at the antipodal singularity and the value of $4\pi\langle \rho+3p\rangle/\Lambda$ are: a) (115.85,6.564) and 1.06, b) (18.05,0.411) and 1.07, c) (9.8,1.37) and 1.01, d) (43.04,4.84) and 1.01. The $\rho(x)$ profile (red curve) is arbitrarily scaled to fit each plot.}
\label{f1}
\end{figure}

When extracting $\alpha$ and $\beta$, deviations due to higher powers of $x_{\rm max}-x$ are also clearly present. We may thus compare our numerical solution for $R(x)$ and $B(x)$ with the series expansions in (\ref{e21}). This vacuum prediction ignores the effects of matter, which we see are most pronounced close to the singularity in the cases of Fig.~1b and 1c. For all four cases, our numerical results confirm the prediction in a suitable range of $x$ from the singularity. For Fig.~1a in particular, the agreement with (\ref{e21}) is within 0.1\% down to $x_{\rm max}-1000$, while the deviation is $\sim$14\% when only the first term is used in each of $R(x)$ and $B(x)$. 

We may instead place a singularity at $x=0$. We then have $\rho(0)=0$ and it is the choice of $\rho_1$ in (\ref{e18}) that determines $\rho(x)$. We choose $4\pi(\rho_1+3w\rho_1)=\Lambda$ for a given $w$. The additional choice of $\alpha$ and $\beta$ for the $x=0$ singularity completes the initial conditions, again specified slightly away from the origin. We give $\alpha$ and $\beta$ on each plot in Fig.~2. Generically, another singularity with different $\tilde\alpha$ and $\tilde\beta$ forms at the antipode. But for a given $\alpha$ we can adjust $\beta$ to obtain a solution that is symmetric about the maximum of $R(x)$, so that $\tilde\alpha$ and $\tilde\beta$ are the same as $\alpha$ and $\beta$. We give three examples of this in Fig.~2a,b,c, where we see a variety of behaviours for different choices of $w$ and $\alpha$. The value of $4\pi\langle \rho+3p\rangle/\Lambda$ in each case is given in the caption.
\begin{figure}[ht]
\centering
\includegraphics[width=\linewidth]{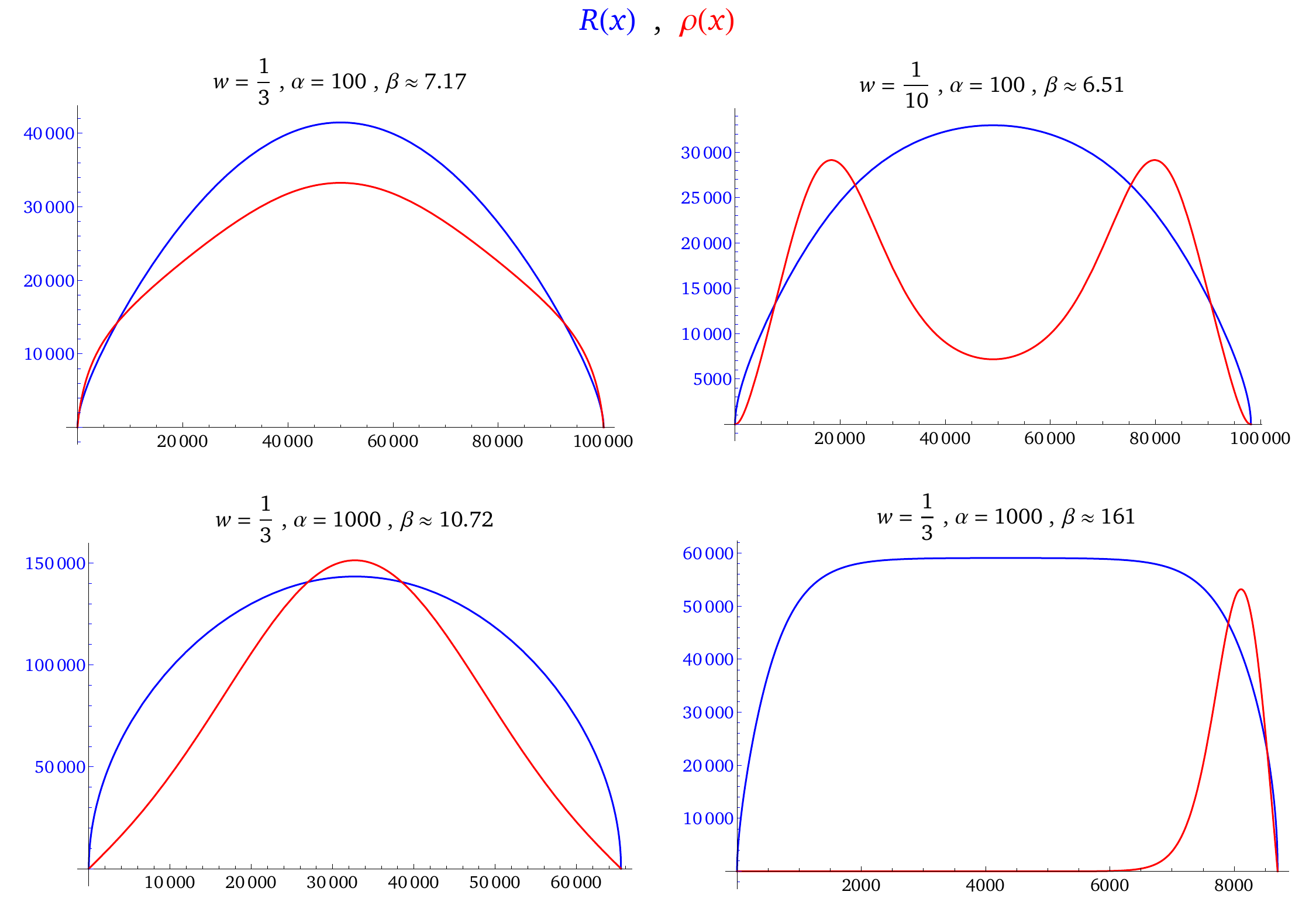}
\caption{The two-singularity solutions with $4\pi(\rho_1+3p_1)=\Lambda$ and with various input values of $w$ and ($\alpha$, $\beta$) at the $x=0$ singularity shown on each plot. The resulting values of $4\pi\langle \rho+3p\rangle/\Lambda$ are: a) 1.064, b) 1.10 , c) 1.87, d) 1000.}
\label{f2}
\end{figure}

A much larger value of this ratio can be achieved in the asymmetric case of two singularities. An example of this is shown in Fig.~2d, where we have increased $\beta$ for the $x=0$ singularity. The antipode singularity then turns out to have $(\tilde\alpha,\tilde\beta)\approx(1746,0.1087)$. This spacetime supports much more matter and the result is $4\pi\langle \rho+3p\rangle/\Lambda\approx1000$; in other words, the singularities now completely dominate the effect of $\Lambda$. Meanwhile, the total Komar mass $M_K^\textrm{total}$ for this spacetime is not dramatically larger than in Fig.~2c.

\section{Stability analysis}\label{s4}
With $B(x)=e^{\nu(x)}$ and $R(x)=e^{\lambda(x)}$ we consider a time-dependent perturbed version of metric (\ref{e0}) where $\nu(x)\to\nu(x)+d\nu(x,t)$ and $\lambda(x)\to\lambda(x)+d\lambda(x,t)$. This choice of metric does not fully fix the gauge, but we shall address this below. In addition, we consider $\rho(x)\to\rho(x)+d\rho(t,x)$ and $p(x)\to p(x)+w\,d\rho(t,x)$, thus preserving $p=w\rho$. We also treat a matter velocity $v(t,x)$ in the $x$ direction as a perturbation, and this introduces off-diagonal components in the energy momentum tensor $\delta T^t_x=\delta T^x_t=v(t,x)(\rho(x)+p(x))$. The Einstein equations at first order in the perturbations can be used to eliminate $d\rho(t,x)$ in favour of $d\nu(t,x)$ and $d\lambda(t,x)$. With $d\nu(t,x)=e^{-i\omega t}d\nu(x)$ and $d\lambda(t,x)=e^{-i\omega t}d\lambda(x)$ we then have the two coupled ODEs.\footnote{The two equations, linear in $d\lambda(x)\to l$ and $d\nu(x)\to n$, are
$w l''+l' \left(3 w \lambda '-(w-1) \nu '\right)+l' \lambda '+(w+1) (l+n) e^{2 \nu -2 \lambda}-(w-1) n' \lambda '+n \left(-\Lambda (w+1) e^{2 \nu }-4 w \lambda ''+4 (w-1) \lambda ' \nu '-2 (3 w+1) \lambda '^2\right)=-\omega^2 l$ and $(w+1) l''+l' \left((3 w+1) \lambda '-(w+1) \nu '\right)+e^{2 \nu } (n (\Lambda-w (\Lambda+16 \pi  \rho ))-2 l (\Lambda-8 \pi  \rho  w))-2 \lambda '' (l+2 w n+n)-2 \lambda '^2 (l+3 w n)+(w-1)(l+n) e^{2 \nu -2 \lambda }-2 (l+n) \nu ''+n''-(w+1) n' \lambda '+4 (w+1) n \lambda ' \nu'=-\omega^2 n$.}
With $v(t,x)=e^{-i\omega t}v(x)$ we also have
\begin{align}
v(x)=-i\frac{\omega}{4\pi}\frac{d\lambda'(x)+d\lambda(x) \left(\lambda '(x)-\nu '(x)\right)-d\nu(x) \lambda '(x)}{e^{2 \nu(x) } (\rho(x)+p(x))}
.\label{e20}\end{align}
$v(x)$ is the magnitude of a physical radial velocity vector and requiring that this vector field remain smooth and continuous implies that $v(x)\to0$ as $R(x)\to0$, that is at the boundaries $x=0$ and $x=x_{\rm max}$. This is true whether or not there are singularities at these points. The metric perturbations $d\nu(x)$ and $d\lambda(x)$ need not vanish at the boundaries.
\begin{figure}[b]
\centering
\includegraphics[width=\linewidth]{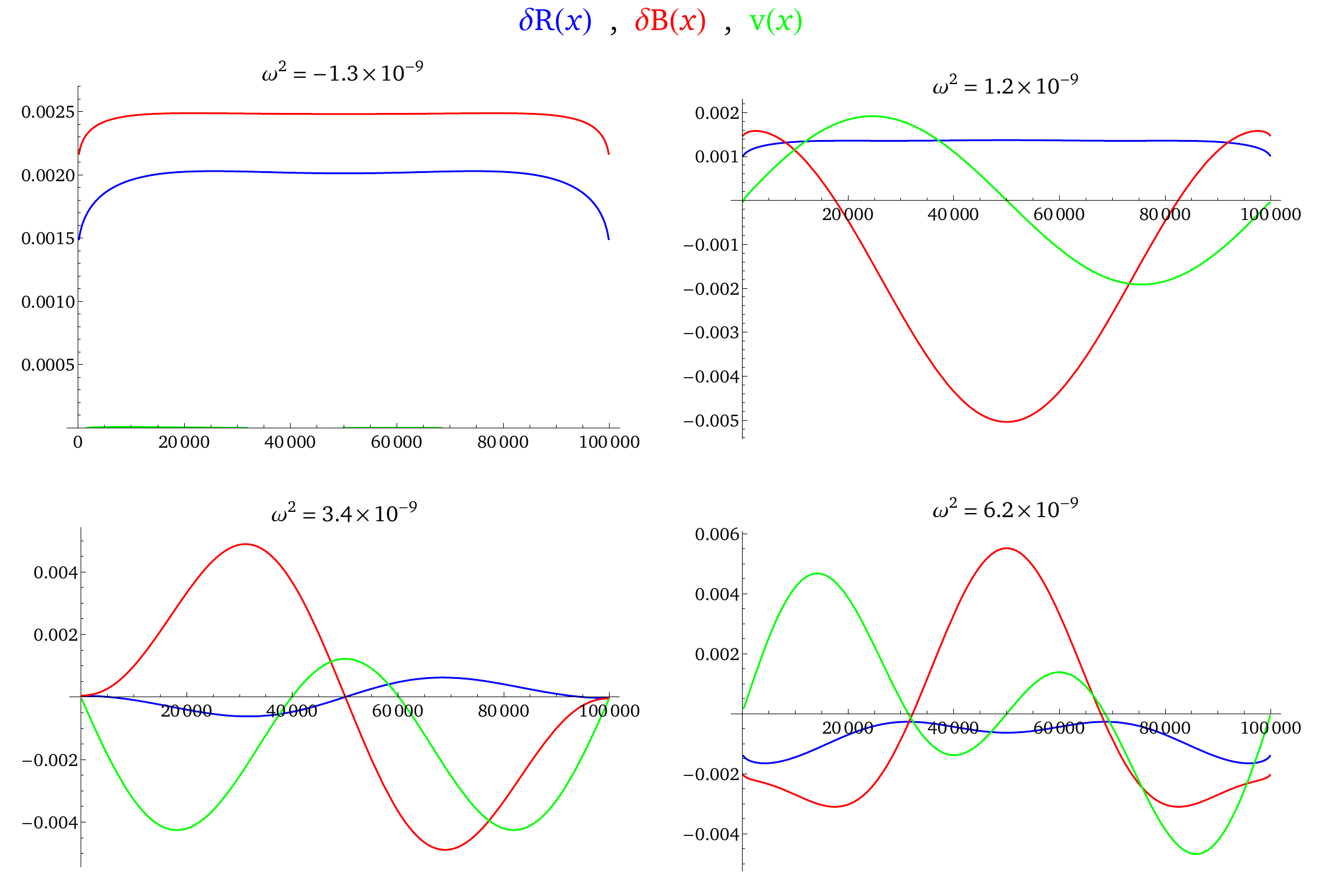}
\caption{The first four $\delta B(x)$ and $\delta R(x)$ eigenfunctions and the velocity profiles, labeled by their eigenfrequencies, corresponding to Fig.~\ref{f2}a. A common overall scale of the velocities (green curves) is adjusted to fit the plots. The velocity of the unstable mode (first plot) is nonzero but small.}
\label{f3}
\end{figure}

We are now able to determine the eigenfunctions $d\nu_n(x)$ and $d\lambda_n(x)$ and the eigenfrequency squares $\omega_n^2$. We find that the $v=0$ boundary conditions ensure that the set of $\omega_n^2$ is real. This indicates that the gravitational perturbation equations have produced a self-adjoint spatial wave operator, with the $v=0$ boundary condition playing a role similar to the finite energy condition for the scalar field equation. This is fortunate since energy is poorly defined in the gravity context. We also find that the metric and matter perturbations remain finite at the boundaries, as is the perturbed scalar curvature.

Some of the eigenfunctions may be pure gauge, and we may eliminate them as follows. By considering a coordinate transformation $x\to x+\xi(x)$ we can obtain the shifts $\delta R(x)$ and $\delta B(x)$. The relation between these shifts is $\delta B(x)=(B(x)\delta R(x)/R'(x))'$. When an eigenfunction satisfies this relation, then it is pure gauge and it should be eliminated. After eliminating the pure gauge modes we numerically recover the known spectrum for the Einstein static universe, $\omega_n^2\propto n(n+2)w-(1+3w)$ for $n=0,2,3,4,...$ \cite{Harrison:1967zza,Gibbons:1987jt,Barrow:2003ni}. Notice that there is a single unstable mode for $w>1/5$, the negative $\omega_0^2$ mode, with more unstable modes developing for $w<1/5$.

The inhomogeneous spacetimes of the last section display spectra with very similar characteristics. For $w=1/3$ we find only one unstable mode for Figs.~1a, 2a, 2c, 2d. The unstable mode yields an overall expansion or contraction, and unlike the ESU, this expansion or contraction mode itself is inhomogeneous. For Figs.~1b, 1c, 1d, 2b we have 3, 3, 2, 2 unstable modes, respectively, with the last three having $w=1/10$.

In Fig.~\ref{f3} we show the first four $\delta B(x)$ and $\delta R(x)$ eigenfunctions and their velocity profiles, for the symmetric case of two singularities shown in Fig.~\ref{f2}a. The only unstable mode appears in Fig.~\ref{f3}a. For this unstable mode, we see that the metric perturbations are close to being uniform except for the bending down near the boundaries, indicating that the singularities have a constraining effect on the evolution. 

\section{Mimicking dark energy (expanding spacetime)}\label{s5}
The discovery that the Einstein static universe has an unstable mode corresponding to the overall expansion or contraction \cite{Edd1} shifted the attention to an expanding universe \cite{Lem,Edd1,Edd2}. We have found that inhomogeneous static solutions have the same stability structure (all modes are stable except for one or a few) despite the presence of singularities. This leads to the question of time-evolving solutions containing singularities and the influence that singularities have on the evolution.

We first mention that there is a scaling law for both one- and two-singularity static solutions. Given a solution $R(x)$, $B(x)$ characterized by quantities ($\Lambda$, $\rho_1$, $R_{\rm max}$, $x_{\rm max}$, $\alpha_i$, $\beta_i$), then there is another solution $R_{\rm new}(x)=\zeta R(\zeta^{-1}x)$, $B_{\rm new}(x)=B(\zeta^{-1}x)$ characterized by
\begin{align}
    (\zeta^{-2}\Lambda,\zeta^{-2}\rho_1,\zeta R_{\rm max},\zeta x_{\rm max},\zeta^\frac{1}{2}\alpha_i,\zeta^\frac{1}{4}\beta_i)
.\end{align}
The ratio $4\pi\langle \rho+3p\rangle/\Lambda=1+\langle(\partial^i B \partial_i B)/B^2\rangle/\Lambda$ is not affected by this scaling; the size of the $B$ gradient term relative to $\Lambda$ remains the same. Both the mass $m_i=\frac{\alpha_i^3}{8\beta_i^2}$ and the Komar mass $M_{Ki}=\frac{1}{4}\alpha_i^2$ scale as $\zeta$, that is, with the spatial size of the spacetime.

Let us introduce the time dependence as $B(x)\to B(t,x)$ and $R(x)\to R(t,x)$ and continue setting off-diagonal components of the metric to zero. We need the corresponding time-dependent Einstein equations, and the analogue of (\ref{e10}) is
\begin{align}
\frac{2}{3}K^2+{}^{(3)}\!R&=16\pi\rho+2\Lambda+2\sigma^2
,\end{align}
where $K= -(1/B) (\dot{B}/B + 2 \dot{R}/R)$ is the trace of the extrinsic curvature and $\sigma^2=(1/3B^2)(\dot{R}/R - \dot{B}/B)^2$ is the shear scalar. To develop the acceleration equation, we can follow \cite{Buchert:1999er} and define a scale factor $a(\tau)$ proportional to the cube root of the proper volume. We use the proper time $\tau$ to define the acceleration $a''(\tau)$. The resulting equation has volume-averaged terms,
\begin{align}
    \frac{3}{a(\tau)}a''(\tau)=-4\pi\langle \rho+3p\rangle +\Lambda+ \langle(\partial^i B \partial_i B)/B^2\rangle -2\langle\sigma^2\rangle +\frac{2}{3}\langle (K-\langle K\rangle)^2\rangle
.\label{e14}\end{align}
The $B$ gradient term arises similarly to the static case.

Unlike the $B$ gradient term, the $\sigma$ and $K$ terms vanish in the static limit. The latter terms appear in \cite{Buchert:1999er}, but as is quite common in such studies, the synchronous gauge with $g_{00}=-1$ and $g_{i0}=g_{0i}=0$ is chosen. Then the $B$ gradient term is absent, and its effect is transferred to other terms. However, trying to apply the synchronous gauge even just to the static metric (\ref{e0}) introduces time dependence and likely global problems.  In any case, for our metric choice, the $B$ gradient term captures physical effects of singularities, as we discussed in Section \ref{s2}, and in particular reflects the finite contribution to the total Komar mass, which is otherwise vanishing for a closed universe. The time dependence of the $B$ gradient term could also reflect the time dependence of the singularities themselves ($\alpha(\tau)$ and $\beta(\tau)$). For an expanding universe, the gradient term acts similarly to a positive $\Lambda$, or a negative pressure, to make the acceleration more positive.

We have explored the inhomogeneous generalization of the Einstein static universe. The possible physical relevance lies in the role the ESU played in the development of homogeneous and isotropic cosmology. The generalization leads to negative mass singularities --- effectively benign, as argued --- and exhibits no further pathological behaviour in the other respects discussed. Our results also suggest that the singularities should be both primordial and large in mass. Being a primordial feature of our universe avoids the difficulty of creating such an object from normal matter (a difficulty suggested by the positive mass theorem \cite{SchoenYau1979,Witten1981}), and being large in mass reduces ``negative mass chasing positive mass'' phenomena \cite{Lutt,Bondi}. Our picture is entirely different from \cite{Farnes:2017gbf,Najera:2021tcx}. In particular, our static solutions point to a cosmologically large mass since $m=\frac{\alpha^3}{8\beta^2}$ is numerically large and since $m$ scales with the size of the universe. Note also that these singularities repel each other \cite{Bondi}. It remains to learn more about cosmologically relevant expanding inhomogeneous solutions, either closed or open, and the effect that some number of singularities could have on the expansion. Would a cosmological constant even be necessary to achieve acceleration?


\begin{thebibliography}{99}
\bibitem{Ein} A.~Einstein, \emph{Kosmologische Betrachtungen zur allgemeinen Relativitatstheorie}, Sitz.~Konig.~Preuss. Akad.~ (1917) 142-152.
\bibitem{ORaifeartaigh:2017uct} C.~O'Raifeartaigh, M.~O'Keeffe, W.~Nahm and S.~Mitton, \emph{Einstein's 1917 Static Model of the Universe: A Centennial Review} Eur.~Phys.~J.~H \textbf{42} (2017) 431-474 arXiv:1701.07261 [physics.hist-ph].
\bibitem{Ishihara:2021gty} H.~Ishihara and S.~Matsuno, \emph{Inhomogeneous generalisation of Einstein{\textquoteright}s static universe with Sasakian space}, PTEP \textbf{2022} (2022) 023E01 [arXiv:2112.02782 [hep-th]].
\bibitem{Belletete:2013nqa} J.~Bellet{\^e}te and M.~B.~Paranjape, ``On negative mass,'' Int. J. Mod. Phys. D \textbf{22}, 1341017 (2013) [arXiv:1304.1566 [gr-qc]].
\bibitem{Mbarek:2014ppa} S.~Mbarek and M.~B.~Paranjape, ``Negative mass bubbles in de Sitter spacetime,'' Phys. Rev. D \textbf{90}, no.10, 101502 (2014) [arXiv:1407.1457 [gr-qc]].
\bibitem{Johnson:2019tgc} M.~C.~Johnson, M.~B.~Paranjape, A.~Savard and N.~Tapia-Arellano, ``Stable, thin wall, negative mass bubbles in de Sitter space-time,'' Gen. Rel. Grav. \textbf{52}, no.8, 80 (2020) [arXiv:1910.01774 [gr-qc]].
\bibitem{HawkingEllis} S.~W.~Hawking and G.~F.~R.~Ellis, \emph{The Large Scale Structure of spacetime} (Cambridge University Press, Cambridge, 1973).
\bibitem{Wald:1984rg} R.~M.~Wald, \emph{General Relativity} (Cambridge University Press, Cambridge, 1984).
\bibitem{Vargas-Serdio:2020qpu} S.~Vargas-Serdio and H.~Quevedo, \emph{Singularity theorems in Schwarzschild spacetimes}, Eur.~Phys.~J.~Plus \textbf{135} (2020) 636 [arXiv:2001.11376 [gr-qc]].
\bibitem{Horowitz:1995gi} G.~T.~Horowitz and D.~Marolf, \emph{Quantum probes of space-time singularities}, Phys.~Rev.~D \textbf{52} (1995) 5670-5675 [arXiv:gr-qc/9504028 [gr-qc]].
\bibitem{Helliwell:2013qxh} T.~M.~Helliwell and D.~A.~Konkowski, \emph{Quantum singularities in spherically symmetric, conformally static spacetimes}, Phys.~Rev.~D \textbf{87} (2013) 104041 [arXiv:1302.3970 [gr-qc]].
\bibitem{Ishibashi:1999vw} A.~Ishibashi and A.~Hosoya, \emph{Who's afraid of naked singularities? Probing timelike singularities with finite energy waves}, Phys.~Rev.~D \textbf{60} (1999) 104028 [arXiv:gr-qc/9907009 [gr-qc]].
\bibitem{Wald:1980jn} R.~M.~Wald, \emph{Dynamics in nonglobally hyperbolic static space-times}, J.~Math.~Phys.~\textbf{21} (1980) 2802-2805.
\bibitem{ReedSimonI} M.~Reed and B.~Simon, \emph{Methods of Modern Mathematical Physics, {I}: Functional Analysis}, Academic Press, New York, 1980.
\bibitem{Ishibashi:2003jd} A.~Ishibashi and R.~M.~Wald, \emph{Dynamics in nonglobally hyperbolic static space-times.~2.~General analysis of prescriptions for dynamics}, Class.~Quant.~Grav.~\textbf{20} (2003) 3815-3826 [arXiv:gr-qc/0305012 [gr-qc]].
\bibitem{Giveon:2004yg} A.~Giveon, B.~Kol, A.~Ori and A.~Sever, \emph{On the resolution of the timelike singularities in Reissner-Nordstrom and negative mass Schwarzschild}, JHEP \textbf{08} (2004) 014 [arXiv:hep-th/0401209 [hep-th]].
\bibitem{Buchert:1999er} T.~Buchert, \emph{On average properties of inhomogeneous fluids in general relativity.~1.~Dust cosmologies},  Gen.~Rel.~Grav.~\textbf{32} (2000) 105-125 arXiv:gr-qc/9906015 [gr-qc].
\bibitem{Harrison:1967zza} E.~R.~Harrison, \emph{Normal Modes of Vibrations of the Universe}, Rev.~Mod.~Phys.~\textbf{39} (1967) 862-882.
\bibitem{Gibbons:1987jt} G.~W.~Gibbons, \emph{The Entropy and Stability of the Universe}, Nucl.~Phys.~B \textbf{292} (1987) 784-792.
\bibitem{Barrow:2003ni} J.~D.~Barrow, G.~F.~R.~Ellis, R.~Maartens and C.~G.~Tsagas, \emph{On the stability of the Einstein static universe}, Class.~Quant.~Grav.~\textbf{20} (2003) L155-L164 [arXiv:gr-qc/0302094 [gr-qc]].
\bibitem{Edd1} A.~S.~Eddington, \emph{On the instability of Einstein’s spherical world.}, MNRAS \textbf{90} (1930) 668-678.
\bibitem{Lem} G.~Lemaitre, \emph{Un univers homogene de masse constante et de rayon croissant, rendant compte de la vitesse radiale des nebuleuses extra-galactiques}, Annal.~Soc.~Sci.~Brux.~Serie A.~\textbf{47} (1927) 49-59.
\bibitem{Edd2} A.~S.~Eddington, \emph{The recession of the extra-galactic nebulae}, MNRAS \textbf{92} (1931) 3-6.
\bibitem{SchoenYau1979} R.~Schoen and S.-T.~Yau, \emph{On the proof of the positive mass conjecture in general relativity}, Commun.~Math.~Phys.~\textbf{65} (1979) 45-76.
\bibitem{Witten1981} E.~Witten, \emph{A new proof of the positive energy theorem}, Commun.~Math.~Phys.~\textbf{80} (1981) 381-402.
\bibitem{Lutt} J.~M.~Luttinger, (1951). \emph{On ``Negative" mass in the theory of gravitation} (PDF). Gravity Research Foundation.
\bibitem{Bondi} H.~Bondi, \emph{Negative Mass in General Relativity}, Rev.~Mod.~Phys.~\textbf{29} (1957) 423-428.
\bibitem{Farnes:2017gbf} J.~S.~Farnes, \emph{A unifying theory of dark energy and dark matter: Negative masses and matter creation within a modified $\Lambda$CDM framework}, Astron. Astrophys. \textbf{620} (2018) A92 [arXiv:1712.07962 [physics.gen-ph]].
\bibitem{Najera:2021tcx} S.~N{\'a}jera, A.~Gamboa, A.~Aguilar-Nieto and C.~Escamilla-Rivera, \emph{On negative mass cosmology in General Relativity}, Astron. Astrophys. \textbf{651} (2021) L13 [arXiv:2105.11041 [gr-qc]].
\end{thebibliography}
\end{document}